# Cyclic Shift Code for SAC-OCDMA Using Fiber Bragg-Grating

Salwa Mostafa, Abd El-Naser A. Mohamed, Fathi E. Abd El-Samie, Ahmed Nabih Zaki Rashed

Department of Electronics and Electrical Communications, Faculty of Electronic Engineering, Menoufia University,

Menouf, 32952, Egypt.

E-mail: salwamostafa@ymail.com,fathi_sayed@yahoo.com, ahmed_733@yahoo.com

*Abstract*

  We proposed a novel code called cyclic shift (CS) code to overcome the drawbacks exists in traditional Spectral Amplitude Coding- Optical Code Division Multiple Access (SAC-OCDMA) codes that have been presented in the past few years. The proposed cyclic shift code has simple construction and large cardinality in selecting the code weight and the number of users. It also has zero cross correlation which allows it to suppress both Multiple Access Interference (MAI) and Phase Induced Intensity Noise (PIIN). Moreover, the frequency bins of the proposed code exist beside each other which reduce the number of filters needed to encode and decode the data. Therefore, the receiver design becomes simple and cost efficient. We compared the performance of our proposed code to the traditional codes and show that our proposed code gives better performance than the traditional SAC-OCDMA codes. A mathematical analysis of CS code has been derived. Simulation analysis for CS code has been carried out using optisystem ver.13.

*Keywords*— Multiple Access Interference (MAI), Phase Induced Intensity Noise (PIIN), Cyclic Shift (CS), Spectral Amplitude Coding (SAC).

I. *Introduction*

   OCDMA has been used widely as one of the spread spectrum techniques that allows multiple users to share the same transmission media, provides asynchronous access to the networks, supports high security to the system, and enhances spectral efficiency. However, many factors effects on the OCDMA performance such as MAI, PIIN, shot noise and thermal noise which degrade its performance and limit the number of users. Excellent code set is one of the essential factors that allows OCDMA to eliminate MAI, suppress PIIN, and provide a simple receiver. Therefore, many researchers aimed to design codes have high autocorrelation, low cross correlation, practical code length, and free cardinality. Encoding schemes also have an important role in enhancing system performance. Among existing schemes SAC showed the best encoding one because it has better transmission performance and simple hardware implementation. Above all, the incoherent system relies on manipulating optical signal amplitude and knowledge of phase information is not needed. Thus, hardware complexity is reduced. As a consequence, we focused on the incoherent system in our research [1].

Several families of codes have been introduced for SAC-OCDMA including an Optical Orthogonal Code (OOC) [2], Modified Quadratic Congruence (MQC) [3], Hadamard, Modified Frequency Hopping (MFH) [4, 5], Random Diagonal (RD) [6], Dynamic Cyclic Shift (DSC) [7], Multi-Service (MS) [8], and Khazani-Syed (KS) [9]code to solve MAI and PIIN problems by designing codes with an ideal in-phase cross correlation $0 \leq \lambda c \leq 1$. Although these codes can reduce MAI, they cannot eliminate PIIN, which degrades system performance and reduces the number of supported users. Moreover, the receivers of these codes are too complex to be able to eliminate interference such as complementary [3], AND [10], Modified-AND [11], and NAND [12] receivers. The purpose of this work is to introduce a novel code able to overcome all the restrictions contained in previous generated codes. In section II, Cyclic Shift (CS) code construction has been introduced. Mathematical analysis of the proposed code has been derived in section III. In section IV, system architecture has been presented. A comparison between the proposed code and the previous codes has been investigated in section V. In section VI, numerical and simulation results have been discussed. Finally, conclusion has been stated in section VII.

**II. Cyclic Shift Code Construction**

Assume that there are two different code sequences $H_i = \{H_1, H_2, \ldots\ldots H_L\}$ and $F_i = \{F_1, F_2, \ldots\ldots F_L\}$. The cross-correlation between the two code sequences can be written as [3]

$$\lambda_C = \sum_{i=1}^{L} H_i F_i \qquad (1)$$

Our proposed CS code for the SAC-OCDMA system can be characterized by *(L, w, K, λc)*, where *L* is the code length, *w* is the weight of the code, and *K* is the number of users. Our proposed code has large cardinality (where it exists for any integer number of users and any weight).

Step (1): Choose the number of users K wanted to support and the code weight required for system

Step (2): Calculate the code length L and the code matrix dimension K × L

$$L = K \times w \qquad (2)$$

Step (3): Determine the positions of ones (P) in the first code sequence by

$$P = \{C_{1i}, ..........C_{1w}\} \qquad (3)$$

*where* $i = \{1,........,w\}$, $C_{1i}$ *is the column number in the first row of code matrix*

Step (4): Complete the construction of the first code sequence by filling the remaining positions by zeroes

Step (5): Obtain the remaining code sequences by cyclic shifting the pervious code sequence bits w bits to the right.

Step (6): Construct the code matrix

**Example: for w=4, K=6**

1- Choose the number of users and the weight K=6 and w=4
2- Calculate the code length L= 6 × 4 =24 and the matrix dimension (6 × 24)
3- Determine the positions of ones in the first code sequence, they are at $\{C_{11}, C_{12}, C_{13}, C_{14}\}$
4- Complete the construction of the first code sequence by filling the remaining positions by zeroes

$$[1\ 1\ 1\ 1\ 0\ 0\ 0\ 0\ 0\ 0\ 0\ 0\ 0\ 0\ 0\ 0\ 0\ 0\ 0\ 0\ 0\ 0\ 0\ 0]$$

5- Obtain the rest code sequences by cyclic shifting the previous code sequence bits w bits to the right

$$\begin{bmatrix} 1 & 1 & 1 & 1 & 0 & 0 & 0 & 0 & 0 & 0 & 0 & 0 & 0 & 0 & 0 & 0 & 0 & 0 & 0 & 0 & 0 & 0 & 0 & 0 \\ 0 & 0 & 0 & 0 & 1 & 1 & 1 & 1 & 0 & 0 & 0 & 0 & 0 & 0 & 0 & 0 & 0 & 0 & 0 & 0 & 0 & 0 & 0 & 0 \end{bmatrix}$$

6- Construct the remaining code sequences to complete code matrix

λc =0

$$\begin{bmatrix} 1 & 1 & 1 & 1 & 0 & 0 & 0 & 0 & 0 & 0 & 0 & 0 & 0 & 0 & 0 & 0 & 0 & 0 & 0 & 0 & 0 & 0 & 0 & 0 \\ 0 & 0 & 0 & 0 & 1 & 1 & 1 & 1 & 0 & 0 & 0 & 0 & 0 & 0 & 0 & 0 & 0 & 0 & 0 & 0 & 0 & 0 & 0 & 0 \\ 0 & 0 & 0 & 0 & 0 & 0 & 0 & 0 & 1 & 1 & 1 & 1 & 0 & 0 & 0 & 0 & 0 & 0 & 0 & 0 & 0 & 0 & 0 & 0 \\ 0 & 0 & 0 & 0 & 0 & 0 & 0 & 0 & 0 & 0 & 0 & 0 & 1 & 1 & 1 & 1 & 0 & 0 & 0 & 0 & 0 & 0 & 0 & 0 \\ 0 & 0 & 0 & 0 & 0 & 0 & 0 & 0 & 0 & 0 & 0 & 0 & 0 & 0 & 0 & 0 & 1 & 1 & 1 & 1 & 0 & 0 & 0 & 0 \\ 0 & 0 & 0 & 0 & 0 & 0 & 0 & 0 & 0 & 0 & 0 & 0 & 0 & 0 & 0 & 0 & 0 & 0 & 0 & 0 & 1 & 1 & 1 & 1 \end{bmatrix}$$

Our proposed code has a zero cross correlation property. Therefore, it suppresses both MAI and PIIN. Furthermore, it has a simple construction, a practical code length, large cardinality and its frequency bins are located beside each other. In OCDMA systems, encoder and decoders utilized by filters and the number of filters needed is equal to the number of frequency bins (number of one's) in code sequence. In our proposed code the frequency bins exist beside each other. So, one filter with large bandwidth can be used instead of representing every frequency bin by a filter. On the other hand, other codes have their frequency bins separated from each other, which increases number of filters needed. As a result, the receiver complexity and the cost for our proposed code is reduced.

### III. Mathematical Analysis of Cyclic Shift (CS) Code

Let $C_K(i)$ indicates the ith item of the Kth CS code sequence. The code properties expressed as [13]

$$\sum_{i=1}^{L} C_K(i) C_N(i) = \begin{cases} w, & \text{for } K = N \\ 0, & \text{else} \end{cases} \tag{4}$$

To simplify our analysis, the following assumptions are used as in [3, 13]

(a) Each light source is perfectly unpolarized and spectrum is flat over the bandwidth $\left[ v_o - \frac{\Delta v}{2}, v_o + \frac{\Delta v}{2} \right]$, where $v_0$ is the central optical frequency and $\Delta v$ is the light source bandwidth (Hz).

(b) The spectral width of each power spectral component is the same.

(c) At the receiver, each user has the same power.

(d) Every bit stream from every user is synchronized.

The Power Spectral Density (PSD) of the received signal can be expressed as [3, 13]

$$r(v) = \frac{P_{sr}}{\Delta v} \sum_{k=1}^{K} d_K \sum_{i=1}^{L} C_K(i)\{u[v-v_o - \frac{\Delta v}{2L}(-L+2i-2)] - u[v-v_o - \frac{\Delta v}{2L}(-L+2i)]\} \quad (5)$$

$P_{sr}$ is the effective received power

$u(v)$ is the unit step function

$$u(v) = \begin{cases} 1, & v \geq 0 \\ 0, & v < 0 \end{cases} \quad (6)$$

The photocurrent expressed as [13]

$$I = R \int_0^\infty G_d(v) dv \quad (7)$$

$G_d(v)$ is the PSD at the photodiode

$$I = R \int_0^\infty \frac{P_{sr}}{\Delta v} \sum_{K=1}^{K} d_K \sum_{i=1}^{L} C_K(i) C_L(i) u\left[\frac{\Delta v}{L}\right] dv = R \frac{P_{sr} w}{L} d_L \quad (8)$$

The noise power can be expressed as [13, 14]

$$\sigma^2 = I_{shot}^2 + I_{thermal}^2 \quad (9)$$

$$\sigma^2 = 2eBI_{dd} + \frac{4K_B T_n B}{R_L} \quad (10)$$

$$I_{dd} = R \int_0^\infty \frac{P_{sr}}{\Delta v} \sum_{K=1}^{K} d_K \sum_{i=1}^{L} C_K(i) C_L(i) u\left[\frac{\Delta v}{L}\right] dv \quad (11)$$

As in [15], when the users are sending bit ''1'',

$$\sum_{i=1}^{K} d_K \approx 1 \tag{12}$$

$$\sigma^2 = 2eBR\frac{P_{sr}w}{L} + \frac{4K_B T_n B}{R_L} \tag{13}$$

When all the users are sending bit ''1" and the probability of transmitting bit '1' at any time for every user is 50% the previous equation became [15]

$$SNR = \frac{I^2}{\sigma^2} = \frac{\left(\frac{RP_{sr}w}{L}\right)^2}{eBR\frac{P_{sr}w}{L} + \frac{4K_B T_n B}{R_L}} \tag{14}$$

Using Gaussian approximation BER is expressed as [3]

$$BER = .5\, erfc\left(\sqrt{SNR/8}\right) \tag{15}$$

Where erfc: a complementary error function, $K_B$: Boltzmann's constant, R: responsivity of the photodiode R = ηe/ $hf_c$.

## IV. System Architecture

On the transmitter side, an encoder encodes the optical pulses from a broadband light source corresponding to the desired user code sequence to generate a unique codeword for every user. Then, an external modulator modulates the data with the generated codewords of each user. After that, the data from different users are combined together and sent over the optical fiber media. On the receiver side, the received signals are spilt to separate the data of each user. Then, data are decoded by the decoder which in our system is the Fiber Bragg-Grading (FBG). After that, photodiode detects the received signal. Finally, a low pass filter is used to remove high frequency noise [14].

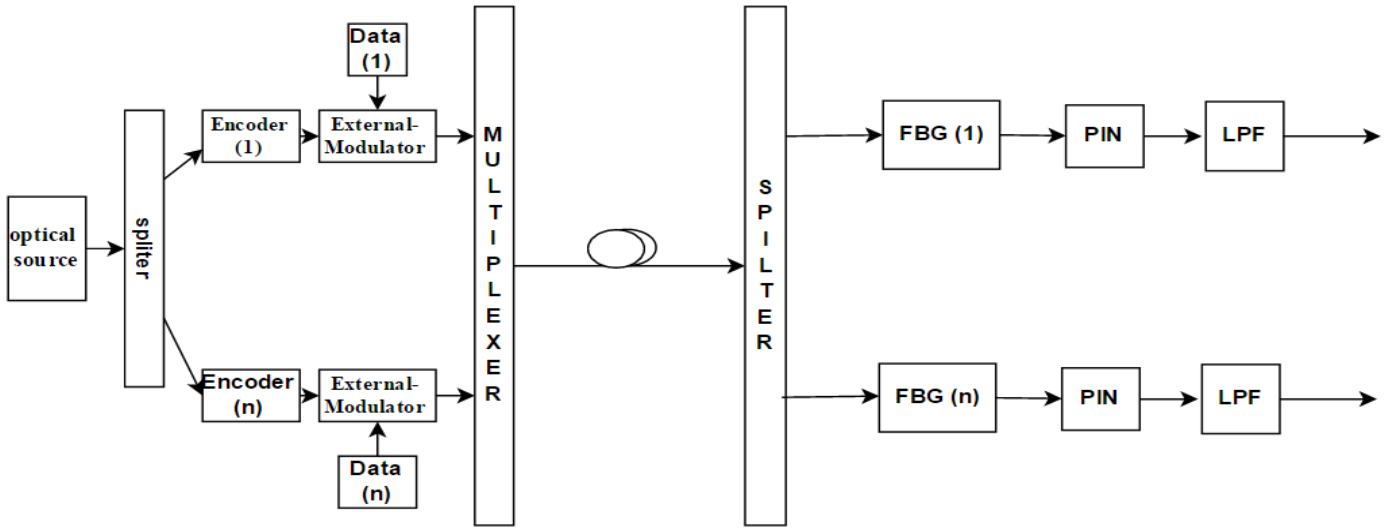

Figure.1 SAC-OCDMA system using CS code with FBG

## V. Codes Comparison

As the code construction properties are one of the most important issues that have an effect on OCDMA system performance. There are some conditions have to be satisfied in any code. These conditions are a minimum cross correlation, the flexibility of choosing the code weight and the number of users supported, and a practical code length. As a consequence, several codes have been constructed in hopes of satisfying these conditions. However, these codes have suffered some restrictions. The following codes (OOC, Hadamard, MQC, MFH, KS, MS, DSC and RD) have an ideal cross correlation $\lambda c=1$, which causes MAI that is produced from overlapping chips between the code sequences. This interference reduces the number of simultaneous users and degrades system performance due to PIIN which is related to MAI. Moreover, OOC has a complicated construction and a very long code length. Hadamard has restrictions in its cardinality. It exists at matrix sequence m only, where $m \geq 2$. Likewise MQC code exists only for prime numbers (p) and $p > 2$, which means minimum weight $w = 4$. Khazani-Syed (KS) code has a simple construction, but the weight is limited to even numbers only. MFH code construction is complex and MS code performance is worse than KS and MFH as its code length is too short. Also, these codes require two branches of decoders at a receiver to decode data and eliminate interferences because of the overlapping between code chips which increase receiver complexity and cost. Furthermore, Single Weight-Zero Cross Correlation (SW-ZCC) [16] has a simple construction and a zero cross correlation, but it has no flexibility in choosing the weight because the weight is always equal to one. The variable weight is needed to support triple-play services. Table II shows the code length needed by the various codes to support 30 users. Finally, Zero Cross Correlation (ZCC) [13] and Multi-Diagonal (MD) [14] codes have a zero cross correlation and the same code length and free cardinality in setting the code weight. They also supported the same number of users as our proposed code. However, our novel code has an advantage over them as its frequency bins occur next to each other which reduce the number of filters needed to encode or decode data. So, our CS code has a simpler and more cost efficient receiver than other codes. Table II presents a comparison between codes in the number of users supported, the code weight, the length, and the cross correlation.

Table I: Comparison between different codes at specific number of users. [16, 17]

| Codes Name | Users Number | Code Weight | Code Length | Cross-correlation |
|---|---|---|---|---|
| OOC | 30 | 4 | 364 | $\leq 1$ |
| Hadamard | 30 | 16 | 32 | 8 |
| MFH | 30 | 7 | 42 | 1 |
| MQC | 30 | 7 | 49 | 1 |
| KS | 30 | 4 | 81 | 1 |
| DSC | 30 | 4 | 30 | $\leq 1$ |
| RD | 30 | 4 | 35 | λc variable in code segment |
| MS | 30 | 4 | 75 | $\leq 1$ |
| SW-ZCC | 30 | 1 | 30 | 0 |
| ZCC code | 30 | 4 | 120 | 0 |
| MD | 30 | 4 | 120 | 0 |
| **CS** | **30** | **4** | **120** | **0** |

Table II: Comparison between different codes

| Codes | Users Number (K) | Code Weight (w) | Code Length (L) | Cross Correlation |
|---|---|---|---|---|
| Hadamard | $K = 2^M - 1$ | $w = 2^{M-1}$ | $L = 2^M$ | $\lambda_C = 2^{M-1}$ |
| MFH | $K = q^2$ | $w = q+1$ | $L = q^2 + q$ | 1 |
| MQC | $K = p^2$ | $w = p+1$ | $L = p^2 + p$ | 1 |
| KS | $K = M\left(\dfrac{w}{2}+1\right)$ | $w = 2, 4, 6, ....$ even number | $L = 3M \sum_{i=1}^{w/2} i$ | 1 |
| DSC | $K = L$ | $w$ | $L = \sum_{i=1}^{w-1} 2^i + D$ | $\leq 1$ |
| RD | $K$ | $w$ | $L = K + 2w - 3$ | $\lambda_C$ variable in code segment |
| MS | $K = M \times w$ | $w$ | $L = M\left(\sum_{i=1}^{w} i - \sum_{i=1}^{w-k_B} i\right)$ | $\leq 1$ |
| SW-ZCC | $K = M^2$ | $w = 1$ | $L = K$ | 0 |
| ZCC | $K = 2^M$ | $w = 2^{M-1}$ | $L = 2^M$ | 0 |
| MD | $K$ | $w$ | $L = K \times w$ | 0 |
| CS | $K$ | $w$ | $L = K \times w$ | 0 |

## VI. Numerical Results and Discussion

The system performance is simulated using MATLAB and the parameters are stated in Table III.

Table III: System Parameters [14]

| broadband source line width | $\Delta v$ | = 3.75 THz | effective power | $P_{sr}$ | = -10 dBm |
|---|---|---|---|---|---|
| operating wavelength | $\lambda$ | = 1550 nm | electrical equivalent noise | B | = 311 MHz |

|  |  |  | band-width of the receiver |  |  |
|---|---|---|---|---|---|
| Receiver noise temperature | $T_n$ | = 300 K | quantum efficiency | $\eta$ | = .6 |
| Planks' constant | h | = 6.626×10$^{-34}$ J.S, | charge of electron | e | = 1.602×10$^{-19}$ C |

Figure 2 shows that at a BER=10$^{-9}$, the CS code supports 90 simultaneous users for w = 4, while KS, MS, DSC and RD support 61, 40, 30 and 50 users respectively at the same weight and the power received ($P_{sr}$ = -10dBm), CS has given better performance because the PIIN is suppressed completely in the CS code due to a zero cross correlation property, while with other codes the PIIN is still present in the system which degrades system performance. Also in Fig.2, MFH and MQC support 60 and 37 simultaneous users at weight 17 and 14 respectively. For MFH and MQC, we have chosen a high weight not w=4 as with other codes, because the number of users supported is limited to the weight w, which is equal to $(w+1)^2$ as indicated in table II. Therefore, CS code can support a large number of users than MFH and MQC with a smaller weight. Hence, CS code gives better performance and supports larger numbers of users than other codes at the same weight and the power received.

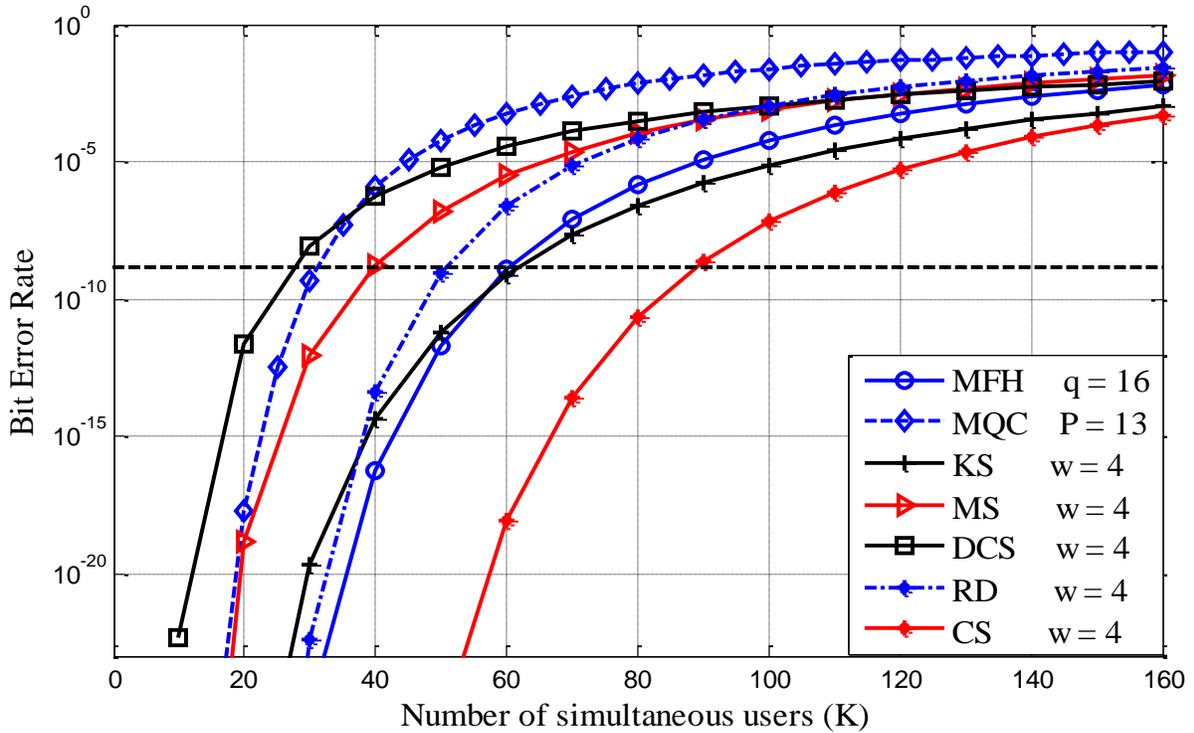

Figure 2. BER versus number of simultaneous users (K) for CS, MFH, MQC, KS, MS, DCS, RD.

Figures 3 shows the BER with the effective power variation. This figure shows that the effective power $P_{sr}$ at the accepted BER of $10^{-9}$ for CS code with (w = 4) is ($P_{sr} \approx$ -12 dBm, ) lower than that required by other codes KS ($P_{sr} \approx$ -10 dBm), RD, DSC, MS and MQC to support the same number of users at the same weight because MAI and PIIN eliminated in our novel code.

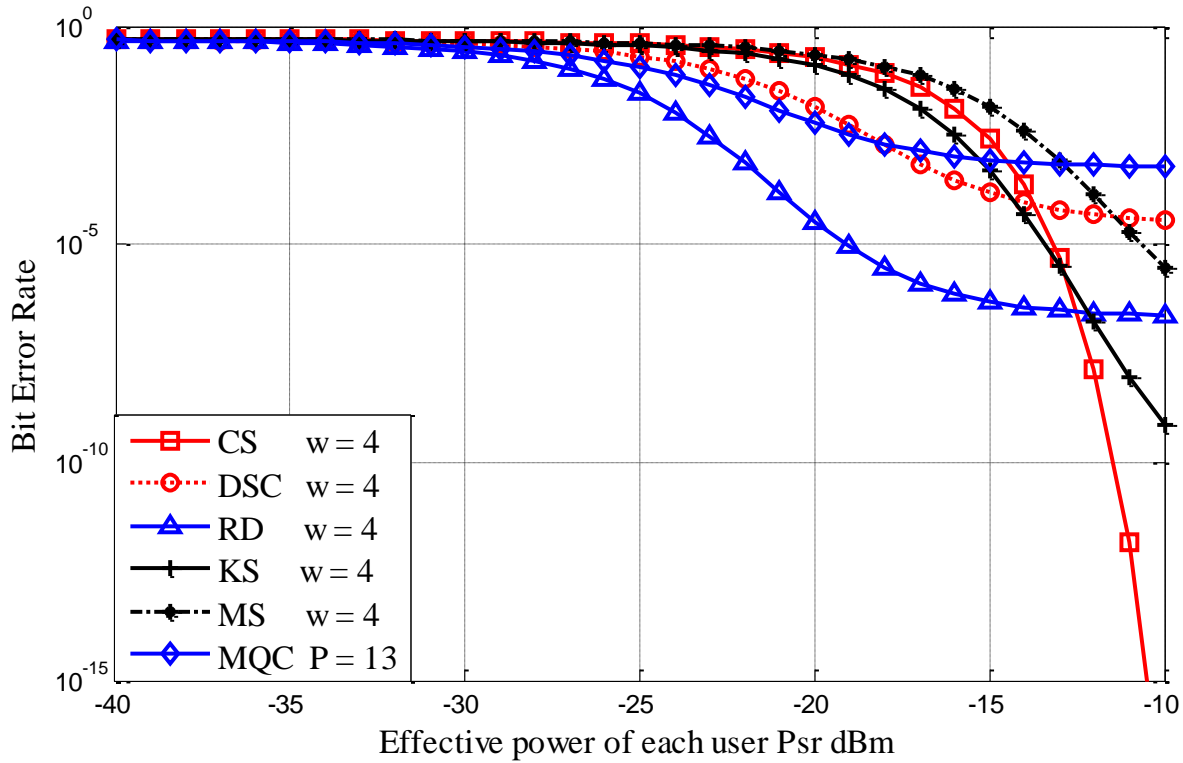

Figure.3 BER Vs. the effective power, when the number of simultaneous users is 60.

## VI.I Network Simulation

Optisystem version.13 is used to simulate the CS code for 4 users with a weight of 4. Figure. 4 shows the simple schematic of CS code in the OCDMA system using the FBG. Table IV indicates the parameters used in the simulation. Nonlinear effects, attenuation, and Group Velocity Dispersion (GVD) have been taken in consideration. The Bit Error Rate (BER) and the eye pattern have been used to characterize the system performance. Figure.5 shows the BER of CS code in the SAC-OCDMA system at different optical fiber distances. It clarifies that as the distance increases, BER increases as well due to the dispersion effect. Then, figure.6 clarifies the performance of the CS code at 40Km, where the BER is equal to $10^{-18}$.

Table. IV the parameters used in optisystem simulation test [14]

| Each chip spectral width | .8 nm | Dispersion | 18 ps/nm-km |
| --- | --- | --- | --- |
| Data rate | 622 Mbps | Dark current | 5 nA |
| Distance (SMF) | 40 Km | Thermal noise coefficient | $1.8\times10^{-23}$ W/Hz |
| Attenuation | .25 dB/km | Low pass filter | .75 GHZ |

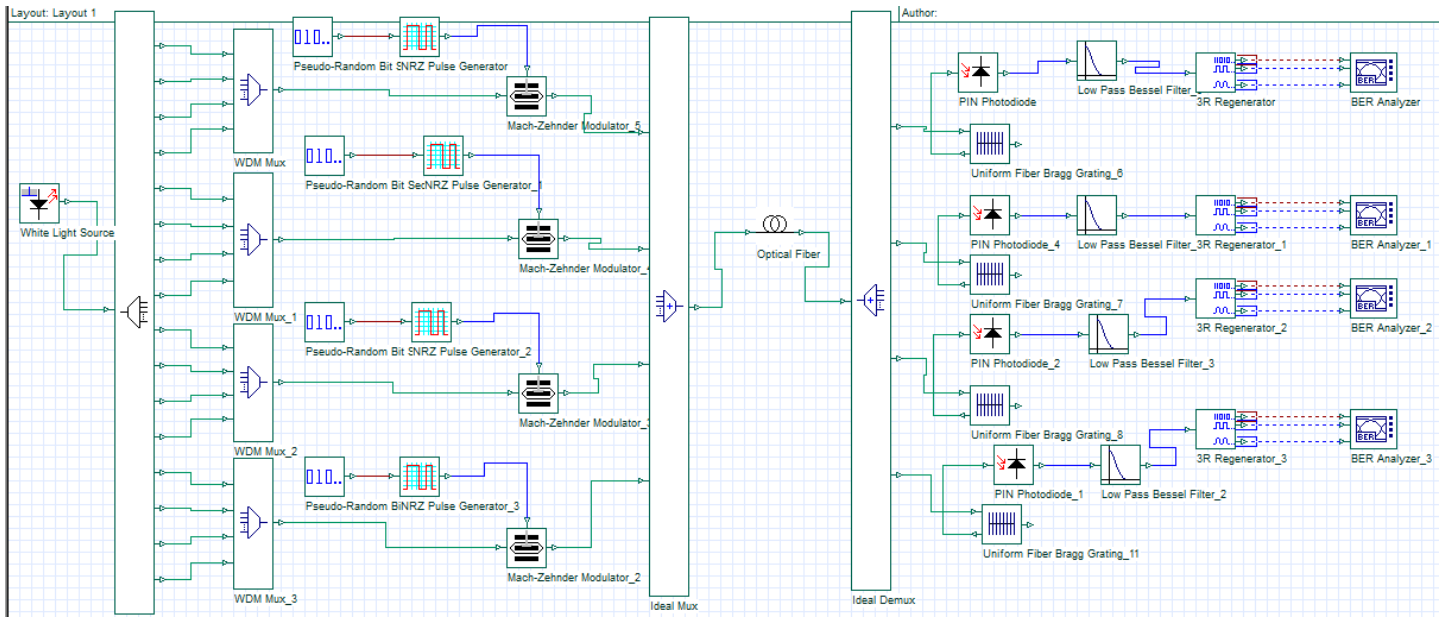

Figure.4 The CS code schematic block diagram for 4 users

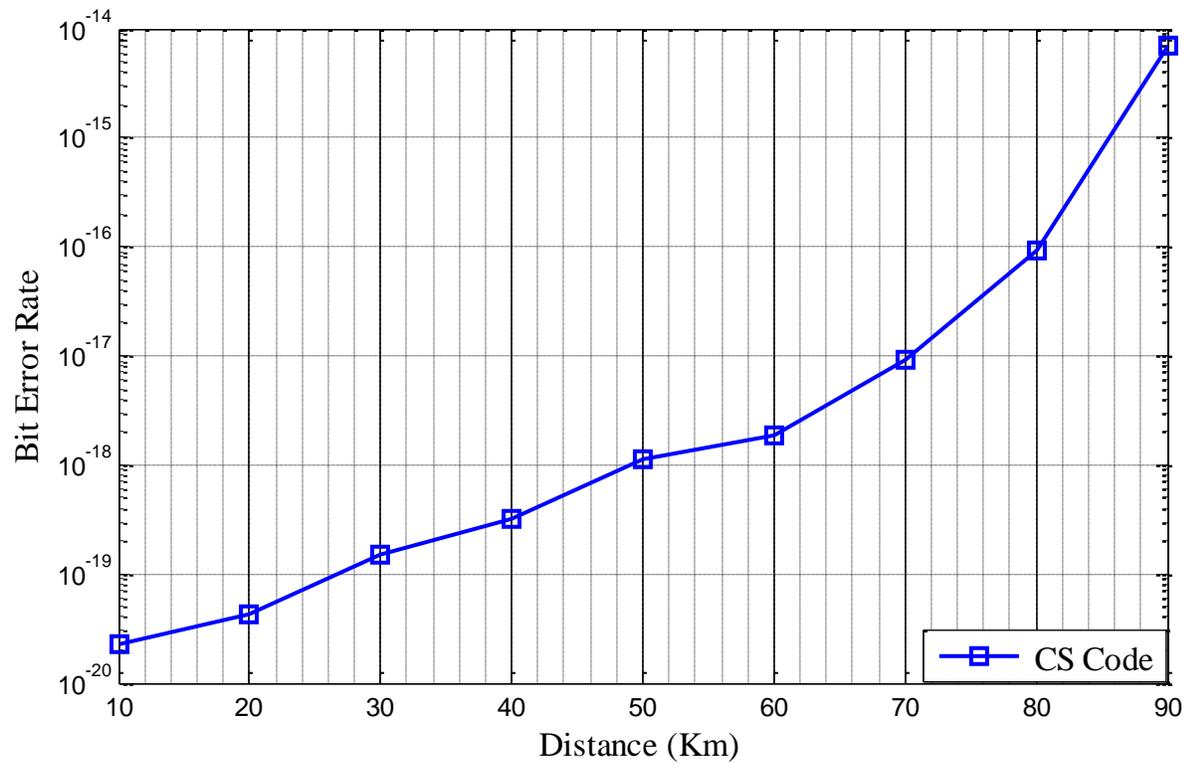

Figure.5 BER Vs. the optical fiber distance

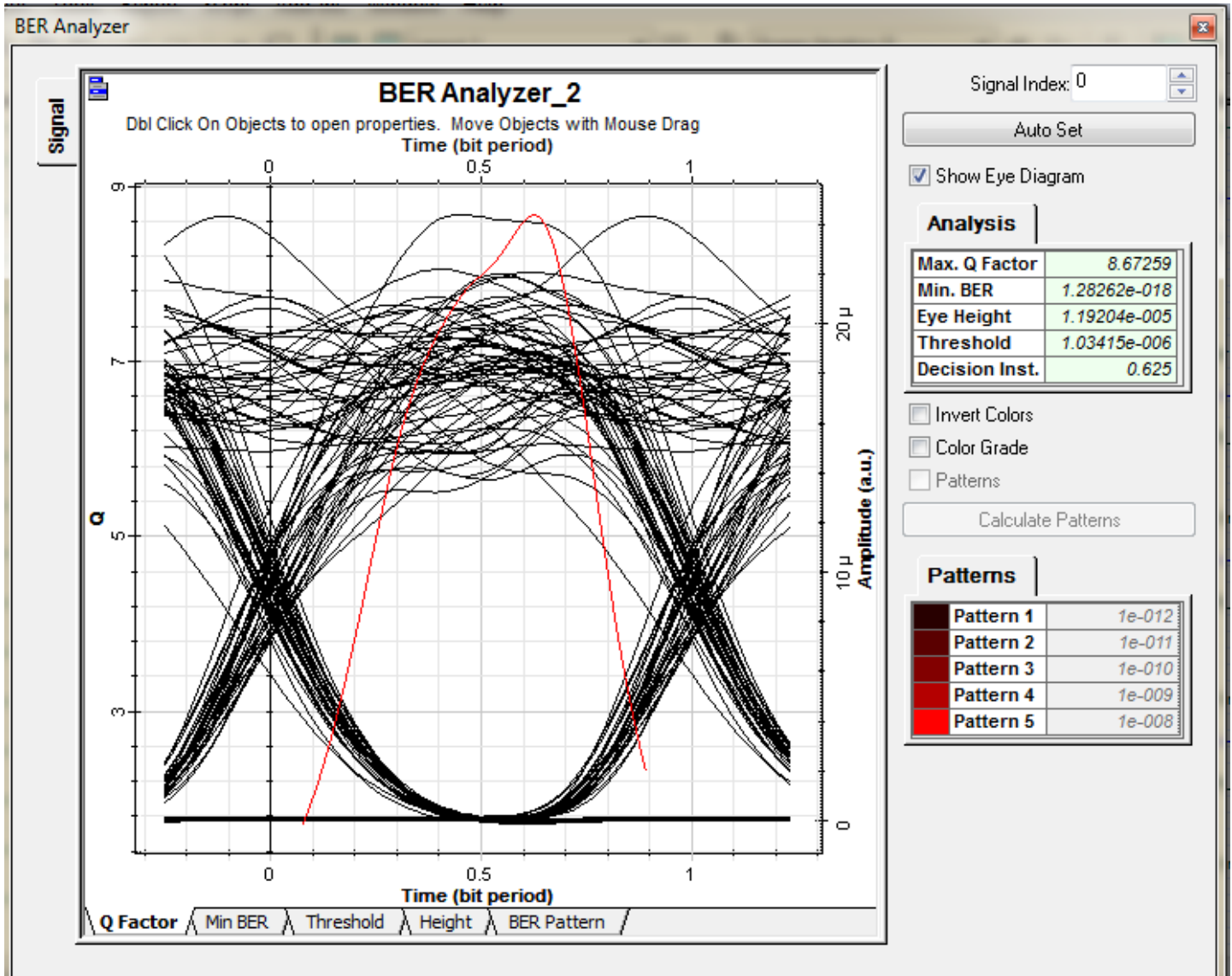

Figure.6 The eye diagram for the CS code at 40Km

## VII. *Conclusion*

The SAC-OCDMA system has been widely used recently. It can suppress the MAI effect, and achieve orthogonality between the code sequences. Codes with a zero cross correlation property, a simple construction, a practical code length, and large cardinality in choosing the code weight and the number of users supported are badly needed. This paper introduced a novel code called Cyclic Shift (CS) code with a zero cross correlation. Therefore, CS code has eliminated both MAI and PIIN. Moreover, our novel code has provided better performance than other codes and has overcome the drawbacks that existed in them. CS code can support a larger number of users with a practical code length and large cardinality in selecting the code weight and the supported users. It also has a simple receiver structure as its frequency bins exist next to each other which reduce the number of filters needed to encode or decode the data.